\documentclass[letterpaper]{jpconf}
\usepackage{graphicx}
\usepackage{amsmath,amssymb,graphicx}
\usepackage[small]{caption2}

\newcommand{\I}{\'\i}
\newcommand{\E}[1]{\ensuremath{\mathrm{E}_{#1}}} 
\newcommand{\SU}[1]{\ensuremath{\mathrm{SU}(#1)}}
\newcommand{\SO}[1]{\ensuremath{\mathrm{SO}(#1)}}
\newcommand{\U}[1]{\ensuremath{\mathrm{U}(#1)}}
\newcommand{\Z}[1]{\ensuremath{\mathbb{Z}_{#1}}} 
\newcommand{\x}{\ensuremath{\times}}

\begin{document}
\title{Moduli fixing in semirealistic string compactifications}

\author{Sa\'ul Ramos-S\'anchez}
\address{Instituto de F\I sica, Universidad Nacional Aut\'onoma de M\'exico, Apdo. Postal 20-364, M\'exico 01000, M\'exico}
\ead{ramos@fisica.unam.mx}

\begin{abstract}
Heterotic orbifold compactifications yield a myriad of models that reproduce many properties
of the supersymmetric extension of the standard model and provide potential solutions to persisting problems
of high energy physics, such as the origin of the neutrino masses and the strong CP problem. However, the 
details of the phenomenology  in these scenarios rely on the assumption of a stable vacuum, characterized by 
moduli fields. In this note, we drop this assumption and address the problem of moduli stabilization in realistic
orbifold models. We study their qualities and their 4D effective action, and discuss how nonperturbative 
effects indeed lift all bulk moduli directions. The resulting vacua, although still unstable, are typically de Sitter and there are generically
some quasi-flat directions which can help to deal with cosmological challenges, such as inflation.

\end{abstract}

\section{Introduction}
There has been a great effort in deriving models of elementary particle physics from string theory. 
Promising models have arisen from intersecting D-branes in type IIB string theory, F-theory,
Calabi-Yau and orbifold compactifications of the heterotic string. In particular, heterotic orbifolds have been 
extensively studied and successful models reproducing several features of the minimal supersymmetric 
extension of the standard model (the MSSM) have been found~\cite{Ibanez:1987sn,Faraggi:1989ka,Kobayashi:2004ya,Buchmuller:2005jr,Buchmuller:2006ik,Lebedev:2007hv,Blaszczyk:2009in}.
Unfortunately, contact between all these different constructions and particle phenomenology requires a full understanding of the details of the vacua that
each model offers. These details are associated with the dynamics of the {\it moduli} fields that describe 
all geometrical features of the particular compactification. For instance, the masses of quarks and leptons 
as well as the dynamics of supersymmetry breakdown depend on the vacuum expectation values (VEVs) of
the moduli, which are arbitrary at tree level in many string constructions. Therefore, in order to achieve
a prediction from string theory and simultaneously to avoid severe cosmological constraints, a mechanism 
that fixes the VEVs of the moduli and gives them large masses is needed. This mechanism is usually called 
{\it moduli fixing} or {\it stabilization}.

Besides constructions that potentially explain particle physics, moduli stabilization has been independently
studied in several scenarios. In type IIB string theory, it has been shown that conjugating fluxes, 
nonperturbative effects and/or $\alpha'$ corrections could yield stable vacua~\cite{Giddings:2001yu,Becker:2002nn,Kachru:2003aw}. 
There has also been some progress in fixing K\"ahler and complex structure moduli in heterotic compactifications
on Calabi-Yau~\cite{Anderson:2010mh} and (generalized) half-flat~\cite{deCarlos:2005kh} manifolds. 
In heterotic orbifold compactifications~\cite{Dixon:1985jw,Dixon:1986jc}, which is the focus of the 
present note, it is frequently argued that the absence of fluxes (other than the NS flux) can render much more difficult --if 
not impossible-- the search for a stable vacuum. However, it is known that the inherent target space modular symmetries~\cite{Ferrara:1989bc,Lauer:1989ax,Lauer:1990tm}
together with nonperturbative effects such as gaugino condensation and string worldsheet instantons  
can provide a scalar potential with metastable minima, at least 
in toy models~\cite{Cvetic:1991qm,Dundee:2010sb,deCarlos:1991gq,deCarlos:1992da,Bailin:1994hu}.

The purpose of the present note is to review the main results of~\cite{Parameswaran:2010ec}.
We go beyond the usual claims and tackle explicitly the problem of moduli fixing
in MSSM candidates arising from heterotic orbifolds. We base our study on a subset of the $\mathcal{O}(300)$ realistic \Z6--II orbifold 
models of the minilandscape~\cite{Lebedev:2006kn,Lebedev:2008un}, in which gaugino condensation can successfully yield an acceptable scale of supersymmetry 
breakdown~\cite{Lebedev:2006tr}. The main phenomenological properties of these models shall be discussed in sec.~\ref{sec:samplemodel}, where
the details of a sample model are given.
We consider all bulk moduli fields, i.e. the dilaton $S$ and those fields describing the deformations in 
size $T_i$ and shape $U_i$ of the compact dimensions. We analyze the 4D effective action of these models, which is well 
understood~\cite{Dixon:1986qv,Hamidi:1986vh,Dixon:1989fj}.
In particular, the scalar potential consists of various computable, perturbative and nonperturbative contributions, which, 
as it turns out, lift all bulk moduli directions. This will be the topic of sec.~\ref{sec:modulifixing}. 
Once moduli stabilization is achieved in these setups, it results immediate to explore some cosmological consequences. 
We speculate in sec.~\ref{sec:cosmology} that inflation might emerge readily, since some of the moduli directions
exhibit very small curvature, rendering the associated fields natural inflaton/curvaton candidates. 

In the following, we shall work in Planck units, specifically we set $M_{Planck}=1$.

\section{An MSSM candidate}
\label{sec:samplemodel}

The so-called minilandscape is perhaps the most fertile search of string-derived MSSM candidates. There have been found
about 300 models with the exact spectrum of the MSSM (or NMSSM~\cite{Lebedev:2009ag}) and no exotics\footnote{The term 
{\it exotic} refers to any field charged under the SM gauge group {\it and} which does not appear in the MSSM matter spectrum.}
at low energies. Besides, 
these models exhibit other appealing properties: approximate gauge coupling unification~\cite{Dundee:2008ts,Anandakrishnan:2011zn}, 
see-saw neutrino masses~\cite{Buchmuller:2007zd}, predominantly TeV gravitino mass~\cite{Lebedev:2006tr}, 
gauge-top unification~\cite{Hosteins:2009xk}, R-symmetries~\cite{Lebedev:2007hv,Lee:2010gv} and other discrete 
symmetries~\cite{Forste:2010pf} that prevent rapid proton decay and other lepton/baryon-number violating processes.

\begin{figure}[t]
\centering
\includegraphics[width=0.8\textwidth]{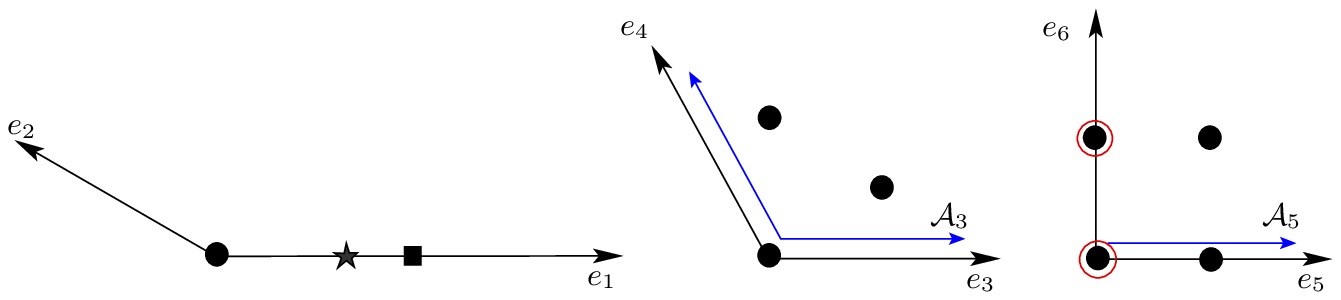}
\caption{\label{fig:Z6IIgeometry} Geometry of the 6D compact space of a \Z6--II orbifold. The 6D space is chosen to be
the torus $T^6 = T^2\x T^2\x T^2$ spanned by the roots $e_i$ of the Lie algebra G${}_2$\x\SU3\x\SO4. The smaller/blue 
arrows denote the noncontractible loops that are embedded as nontrivial Wilson lines $\mathcal{A}_i$ in the gauge degrees 
of freedom. Modding out a \Z6 symmetry in this geometry leaves the points $\bullet$, $\star$ and $\blacksquare$ untouched.}
\end{figure}

The minilandscape is based on \Z6--II orbifold compactifications of the \E8\x\E8 heterotic string. The 6D torus is chosen
to be the product of three 2-tori, $T^6 = T^2\x T^2\x T^2$, which are spanned by the simple roots of the Lie 
algebra G${}_2$\x\SU3\x\SO4, as depicted in fig.~\ref{fig:Z6IIgeometry}. 
In this basis, the action of the \Z6--II symmetry acts identifying the points related by a rotation (or repeated rotations) by 
$\pi/3$ in the G$_2$ 2-torus, by $2\pi/3$ in the \SU3 2-torus, and reflection(s) through the origin in the
last 2-torus. The special points that are left unaffected by this identification, depicted by
$\bullet$, $\star$ and $\blacksquare$ in fig.~\ref{fig:Z6IIgeometry}, are called fixed points and are of crucial 
relevance in the minilandscape, as we shall see shortly. The intrinsic modular invariance of the heterotic string
demands the orbifold action to be embedded in the gauge degrees of freedom, producing thereby the breakdown
of the original \E8\x\E8 gauge group to a subgroup thereof, which in the minilandscape models takes the generic form 
$\SU3_c\x\SU2_L\x\U1_Y\x G_{hidden}\x\U1^n$. All standard model (SM) fields are uncharged under the nonabelian gauge
factor $G_{hidden}$, but some other ``hidden" fields form nontrivial representations under this group.

The guiding principle of the minilandscape search is local grand unification~\cite{Kobayashi:2004ya,Buchmuller:2005jr,Forste:2004ie}, 
which refers to the possibility that certain fixed points
of an orbifold be endowed with the gauge symmetry of a grand unified theory (GUT) like \SU5 or \SO{10}, while the
4D gauge group be the one of the SM. This indeed happens in the small moduli-space region of the minilandscape
and leads to scenarios that preserve the appealing features 
of GUTs such as gauge coupling unification while solving their common phenomenological drawbacks  
such as problematic quark/lepton mass relations. The crucial advantage is that matter fields living at
points endowed e.g. with an \SO{10} local symmetry can transform as {\bf 16}-plets locally. Since the {\bf 16} spinor of 
\SO{10} contains a SM family of quarks and leptons, if ideally there were three of such special local GUTs, one would 
find a geometric origin of the three SM generations. Moreover, the SM Higgs doublets need not arise from such local GUTs,
but from the bulk or a fixed point with no local GUT. Therefore, no doublet-triplet splitting must be enforced in this 
situation. Unfortunately, the scenario with three degenerate SM families coming from local GUTs is not favored
by orbifold constructions~\cite{Lebedev:2006kn}. Instead, only two generations arise from equivalent local \SO{10} GUTs 
(see the encircled/red $\bullet$ in fig.~\ref{fig:Z6IIgeometry}), what is not necessarily bad news, for we might prefer 
a setting in which a distinction between the two lightest and the heaviest SM generation emerges naturally.

\begin{center}
\begin{table}[t]
\centering
\caption{\label{tab:spectrum}
The matter spectrum of a realistic \Z6--II orbifold model. It corresponds to the spectrum of the MSSM plus
standard model singlets and vectorlike exotics, which acquire masses through a Higgs-like effect after some singlets
attain VEVs. The bulk moduli are uncharged under the gauge group. Quantum numbers w.r.t. the gauge group
$\SU3_c\x\SU2_L\x[\SO8\x\SU3]_{hidden}$ are given in parenthesis. The hypercharge $\U1_Y$ appears as subindex.} 
{\small
\begin{tabular}{|cll||cll|}
 \hline
 \multicolumn{6}{|c|}{{Bulk moduli:} $S,\, T_1,\, T_2,\, T_3,\, U$}\\
 \hline
 \multicolumn{6}{|c|}{3 (net) generations}\\
 \hline
 3 & $\left(\boldsymbol{3},\boldsymbol{2};\boldsymbol{1},\boldsymbol{1}\right)_{1/6}$  & $q_i$
 & & &
 \\
 3 & $\left(\overline{\boldsymbol{3}},\boldsymbol{1};\boldsymbol{1},\boldsymbol{1}\right)_{-2/3}$ & $\bar u_i$
 & & &
 \\
 3 & $\left(\boldsymbol{1},\boldsymbol{1};\boldsymbol{1},\boldsymbol{1}\right)_{1}$ & $\bar e_i$  
 & & &
 \\
 3+4 & $\left(\overline{\boldsymbol{3}},\boldsymbol{1};\boldsymbol{1},\boldsymbol{1}\right)_{1/3}$ & $\bar d_i$
 & 
 4 & $\left(\boldsymbol{3},\boldsymbol{1};\boldsymbol{1},\boldsymbol{1}\right)_{-1/3}$  & $d_i$
 \\
 3+4 & $\left(\boldsymbol{1},\boldsymbol{2};\boldsymbol{1},\boldsymbol{1}\right)_{-1/2}$  & $\ell_i$
 &
 4 & $\left(\boldsymbol{1},\boldsymbol{2};\boldsymbol{1},\boldsymbol{1}\right)_{1/2}$  & $\bar \ell_i$
 \\
 \hline
 \multicolumn{6}{|c|}{Higgses}\\
 \hline
 1 & $\left(\boldsymbol{1},\boldsymbol{2};\boldsymbol{1},\boldsymbol{1}\right)_{-1/2}$  & $h_d$
 &
 1 & $\left(\boldsymbol{1},\boldsymbol{2};\boldsymbol{1},\boldsymbol{1}\right)_{1/2}$  & $h_u$
 \\
 \hline
 \multicolumn{6}{|c|}{Standard model singlets}\\
 \hline
 43 & $\left(\boldsymbol{1},\boldsymbol{1};\boldsymbol{1},\boldsymbol{1}\right)_{0}$ & $n_i$
 &
 8 & $\left(\boldsymbol{1},\boldsymbol{1};\boldsymbol{8},\boldsymbol{1}\right)_{0}$ & $\widetilde{N}_i$
 \\
 10 & $\left(\boldsymbol{1},\boldsymbol{1};\boldsymbol{1},\boldsymbol{3}\right)_{0}$ & $N_i$
 &
 10 & $\left(\boldsymbol{1},\boldsymbol{1};\boldsymbol{1},\overline{\boldsymbol{3}}\right)_{0}$ & $\bar{N}_i$
 \\
 \hline
  \multicolumn{6}{|c|}{Exotics}\\
 \hline
  8 & $\left(\boldsymbol{3},\boldsymbol{1};\boldsymbol{1},\boldsymbol{1}\right)_{1/6}$ & $\delta_i\phantom{1^{1^1}}$
 &  
 8 & $\left(\overline{\boldsymbol{3}},\boldsymbol{1};\boldsymbol{1},\boldsymbol{1}\right)_{-1/6}$ & $\bar\delta_i$
 \\
 8 & $\left(\boldsymbol{1},\boldsymbol{1};\boldsymbol{1},\boldsymbol{1}\right)_{1/2}$ & $s^+_i$
 &
 8 & $\left(\boldsymbol{1},\boldsymbol{1};\boldsymbol{1},\boldsymbol{1}\right)_{-1/2}$ & $s^-_i$
 \\
 16 & $\left(\boldsymbol{1},\boldsymbol{2};\boldsymbol{1},\boldsymbol{1}\right)_{0}$ & $m_i$
 & & &
  \\
 \hline
\end{tabular}
}
\end{table}
\end{center}
\vskip -11mm

Let us focus now on one particular minilandscape model. The resulting matter spectrum is presented in tab.~\ref{tab:spectrum}.
The 4D gauge group is $\SU3_c\x\SU2_L\x\U1_Y\x[\SO8\x\SU3]_{hidden}\x\U1^6$. Note that no SM field is charged under 
$[\SO8\x\SU3]_{hidden}$, what ``hides" this sector in the sense that it only interacts gravitationally with the observed matter.
In addition to the 4D $\mathcal{N}=1$ supergravity multiplet (which is not displayed in tab.~\ref{tab:spectrum}),
one obtains a net number of three SM generations, the up and down Higgses required in the MSSM, some
$\SU3_c\x\SU2_L\x\U1_Y$ singlets and a bunch of vectorlike exotics that are decoupled\footnote{It has been checked that e.g. 
operators $n_i^x\ \ell_j\ \bar\ell_k$ exist, implying that the 
additional four (exotic) pairs $\ell_i - \bar \ell_i$ attain large masses once 
some $n_i$ develop VEVs preserving $\mathcal{N}=1$ supersymmetry. All other exotic fields 
($\bar d_i, d_i, \bar \delta_i, \delta_i, s^\pm_i$ and $m_i$) are decoupled analogously, leaving only three massless SM families.} 
at a scale $M_d$ without breaking supersymmetry, but producing the spontaneous breakdown of the additional $\U1^6$.
The SM singlets charged 
under the hidden group can also acquire masses of order $M_d$,  
leaving a pure Yang-Mills \SO8\x\SU3 hidden sector at lower energies. This situation leads to gaugino condensation
in this sector, which renders the scale of supersymmetry breakdown as low as $\sim100$ TeV~\cite{Lebedev:2006tr}, 
not far from the admissible/expected value.

In the absence of matter parity in this model\footnote{We point out that there are many other models which
do exhibit matter parity~\cite{Lebedev:2007hv},\cite[app. E]{RamosSanchez:2008tn}.}, 
some of the SM singlets (those without VEVs) can be considered right-handed neutrinos.
We have verified that they become massive when the exotics decouple, i.e. they get masses of order $M_d\sim M_{GUT}$
too. This large right-handed neutrino mass enables the see-saw mechanism to produce left-handed
neutrino masses of order $10^{-3}-10^{-2}$ eV, as suggested by neutrino-oscillation probes~\cite{Schwetz:2008er}.

As shown in tab.~\ref{tab:spectrum}, this model has five bulk moduli. The dilaton $S$ arising from the supergravity multiplet,
and four deformation parameters. The latter are obtained from demanding invariance of the metric 
$g_{\alpha\beta}=e_\alpha\cdot e_\beta$ under the \Z6--II orbifold action. The free parameters in this case are the ``radius'' of
the cycle $e_1$ in the G$_2$ torus, $e_3$ in the \SU3 torus, and the cycle radii and angle of the last torus, i.e. the magnitude of 
$e_5$ and $e_6$ and the angle between them. These deformations are usually collected in three K\"ahler moduli $T_1,T_2,T_3$ for
the radii of each of the tori, and a complex structure modulus $U$ for the angle in the last torus. 

As a last remark, we would like to mention that this model possesses some useful symmetries. Like many minilandscape 
models, it presents an \SO2 family symmetry~\cite{Raby:2011xx}\footnote{This symmetry is preserved up to order 9 in 
the superpotential, but must be broken to a discrete subgroup at higher orders, for there are no continuous symmetries other 
than gauge symmetries in string constructions.} which is the result of the 
vertical symmetry between the fixed points of the last torus (see e.g. the encircled/red $\bullet$ in fig.~\ref{fig:Z6IIgeometry}).
This symmetry is particularly useful for particle phenomenology, as it could shed light on the structure of the quark
and lepton mass matrices.
Besides, all orbifold models exhibit target-space modular symmetries~\cite{Ferrara:1989qb,Ibanez:1992hc} that act as 
discrete transformations on the moduli, winding numbers and momenta. In the present model, they are
$[\text{SL}(2,\Z{})\x\Gamma_1(3)\x\Gamma_0(4)]_{T}\x[\Gamma^0(4)]_U$ transforming the three
K\"ahler $T_i$ and one complex structure $U$ moduli, respectively.\footnote{Details on these symmetries can be
found in~\cite[app. A.4]{Parameswaran:2010ec}.} These discrete transformations do not only ensure (and help to verify) 
the consistency of the theory, but are a great tool for moduli fixing, as we shall see.

\section{Fixing the moduli}
\label{sec:modulifixing}

In order to stabilize the moduli, a full knowledge of the effective supergravity theory is required. 
Remarkably, heterotic orbifold compactifications are the only scenario in which all relevant 
quantities can be computed explicitly from string theory. Perturbatively, the effective theory
receives contributions from nonvanishing couplings between matter states that can be found by applying 
the so-called selection rules~\cite{Dixon:1986qv,Hamidi:1986vh},
based on the string symmetries that are left unbroken by the orbifold action.
The coupling strengths have been worked out using conformal field theory 
techniques~\cite{Burwick:1990tu,Erler:1992gt,Choi:2007nb}
finding that they are exponentially suppressed by the K\"ahler moduli. Perturbative $\alpha'$ corrections
do not contribute to the tree-level superpotential~\cite{Dine:1985kv}.
At nonperturbative level, the main contribution to the effective theory comes from gaugino condensation,
which is affected by threshold corrections due to the decoupling of exotic matter and string 
excitations~\cite{Dixon:1990pc,Mayr:1993mq}.

To simplify somewhat our discussion, we adopt the following assumptions:
\begin{itemize}
\item all (vectorlike) exotics decouple consistently with supersymmetry at a unique scale $M_d$, 
      such that $(M_{GUT}\sim)M_d\lesssim M_{string}\lesssim 1$;\footnote{In fact,
      this assumption has been confirmed in several cases~\cite{Lebedev:2007hv,Lee:2010gv}.}
\item at some scale close to $M_d$, the charged hidden matter $\widetilde{N}_i,N,\bar{N}$
      are decoupled too, yielding hidden-sector gaugino condensation at the scale $\Lambda_{gc}\lesssim M_d$;
\item nonrenormalizable couplings among matter fields are negligible;
\item the overall volume is $\gg 1$, implying $T_i>1$;
\item some SM singlets (those present in admissible renormalizable couplings) $n_i$ do not enter in the process of 
      decoupling of unwanted matter and attain VEVs $A_i\ll 1$. It follows that matter kinetic terms
      proportional to $|A_i|^2$ in the K\"ahler potential can be safely ignored;	
\item no twisted moduli nor blow-up modes appear, so that we can remain at the orbifold point;
\item D-term contributions to the scalar potential are negligible.
\end{itemize}

Under these conditions, the K\"ahler potential (in the large-volume limit) of the model introduced in sec.~\ref{sec:samplemodel} 
is computable~\cite{Dixon:1989fj,Ibanez:1992hc,Lust:1991yi} and given by
\begin{equation}
\label{eq:K}
  K = -\log\left[S+\bar S - \frac{19}{24\pi^2} \log(T_1 + \bar T_1)
                            + \frac{7}{24\pi^2} \log(T_2 + \bar T_2)  \right] 
        -\sum_j\log(\phi_j+\bar \phi_j)\,,
\end{equation}
where we have included 1-loop effects and denoted all bulk moduli by $S$ and $\phi_j=\{T_1,T_2,T_3,U\}$.

The superpotential is split in the perturbative bit coming from matter couplings $W_{yuk}$ and the gaugino 
condensation part $W_{gc}$; thus, for our model we have
\begin{subequations}
\label{eq:W}
\begin{eqnarray} 
 W_{yuk}&=& 2 \, N_{255}\, A_1^2 \, e^{-2\pi \, T_2/3} \left(A_2 + A_3 e^{-2\pi \, T_1/3}\right)+\ldots\,,\\
 W_{gc} &=& -\frac{c}{e}\,\frac{3}{16 \pi^2} \,\, e^{-4\pi^2 S/3} \, M_d^{3/2}\,        
    \eta(3\,T_2)^{-16/9} \, \left[ \eta(4\,T_3)\,\eta(U/4)\right]^{1/3} \\
 && 
  -\frac{c}{e}\,\frac{3}{32 \pi^2} \, e^{-8\pi^2 S/3} \, M_d^{10/3}\,        
    \eta(3\,T_2)^{4/9} \,\,  \left[ \eta(4\,T_3)\,\eta(U/4)\right]^{2} \,,\nonumber
\end{eqnarray}
\end{subequations}
where $c$ is an unknown constant arising from integrating out the condensate, $\eta(p_i \phi_i)$ is the well-known Dedekind function,
and $N_{255}\approx1.6$ comes from the explicit computation of string worldsheet instantons. 
The two contributions to $W_{gc}$ come from the two hidden groups, \SO8 and \SU3 respectively.
Two remarks are in order:
1) the ellipsis in $W_{yuk}$ contains terms with higher powers in $e^{-T_i}$ which we neglect because all $T_i$ are 
assumed to be large; and 2) modular symmetries control
the moduli dependence of the eta functions, e.g. only $\eta(4\,T_3)$ transforms covariantly under the unbroken
$\Gamma_0(4)$ modular symmetry. 

The structure of the superpotential is very rich. 
First, $W(S)$ takes a racetrack form which can indeed lead to dilaton stabilization~\cite{deCarlos:1991gq,deCarlos:1992da}
(although it must be noticed that the \SU3 contribution is subleading mainly due to the larger power of $M_d$).
Second, $W(T_1)$ is KKLT-like which is known to yield potentially stable vacua with no or little fine-tuning.
Third, $W(T_2,T_3,U)$ is a complicated combination of cusp forms, which exhibits
several critical points. We would like to point out that the inclusion of matter fields in $W$ could be crucial
for arriving to a de Sitter vacuum~\cite{Lebedev:2006qq}. It is also noticeable that, contrary to what happens in
toy models in which there are gazillions of tunable parameters, the only parameters left free to tune in
this promising model are 
$A_i,M_d$ and $c$, connected to the decoupling of unwanted states and the gaugino condensates. 
Therefore, the next task is just to compute the scalar potential and locate a minimum
by varying these few free parameters around admissible values: $c\sim 1,\,A_i\sim M_d\ll 1$. 

\begin{center}
\begin{table}[t]
\centering
\caption{\label{tab:unstabledSI}An unstable de Sitter solution of an orbifold MSSM candidate. 
The dominant contribution to the 
mass eigenstate $\rho_i$ arises from the modulus $\Phi_i\in\{T_1,T_2,T_3,U, S\}$. 
We give the solution to 5 significant figures, but we 
have computed it to a precision of 1000. The value of $\Lambda$ is given in Planck units.
The tachyonic eigenstate is given by ${\rm Re}\rho_2\sim0.9{\rm Re}T_2+0.4{\rm Re}S$.} 
\vskip 2mm
{\small
\begin{tabular}{|c|c|c|}
\hline
\multicolumn{3}{|c|}{Parameters}\\
$c=1/10$ & $20A_1=100A_2=A_3=1/10$ & $M_d=1/65$ \\ 
\hline
\hline
\multicolumn{3}{|c|}{Moduli VEVs}\\
$\Phi_i$ & ${\rm Re}\langle\Phi_i\rangle$ & ${\rm Im}\langle\Phi_i\rangle$ \\
\hline
$T_1$ & $3.51166$ &  $3/2$ \\ 
\hline
$T_2$ & $0.24201$ &  $-1/3$ \\
\hline
$T_3$ & $7.05981$ &  $35/8$ \\
\hline
$U$ & $112.95695$ &  $-506$ \\
\hline
$S$ & $0.09385$ &  $1079/144\pi$ \\
\hline
\end{tabular}
\begin{tabular}{|c|c|c|}
\hline
\multicolumn{3}{|c|}{}\\
\multicolumn{3}{|c|}{Cosmological constant $\Lambda = 6.45\x10^{-19}$}\\
\hline
\hline
\multicolumn{3}{|c|}{Mass eigenstates}\\
$\rho_i\sim \Phi_i$ & $m^2_{{\rm Re}\rho_i}$ & $m^2_{{\rm Im}\rho_i}$ \\
\hline
$\rho_1$ & $1.87\x10^{-18}$ & $3.57\x10^{-18}$ \\
\hline
$\rho_2$ & $-4.76\x10^{-17}$ & $1.2\x10^{-17}$ \\
\hline
$\rho_3$ & $3.15\x10^{-20}$ & $3.29\x10^{-33}$ \\
\hline
$\rho_U$ & $9.84\x10^{-23}$ & $1.51\x10^{-94}$ \\
\hline
$\rho_S$ & $1.14\x10^{-16}$ & $2.16\x10^{-16}$ \\
\hline 
\end{tabular}
}
\end{table}
\end{center}
\vskip -10mm

\begin{figure}[t]
\centering
\includegraphics[width=0.8\textwidth]{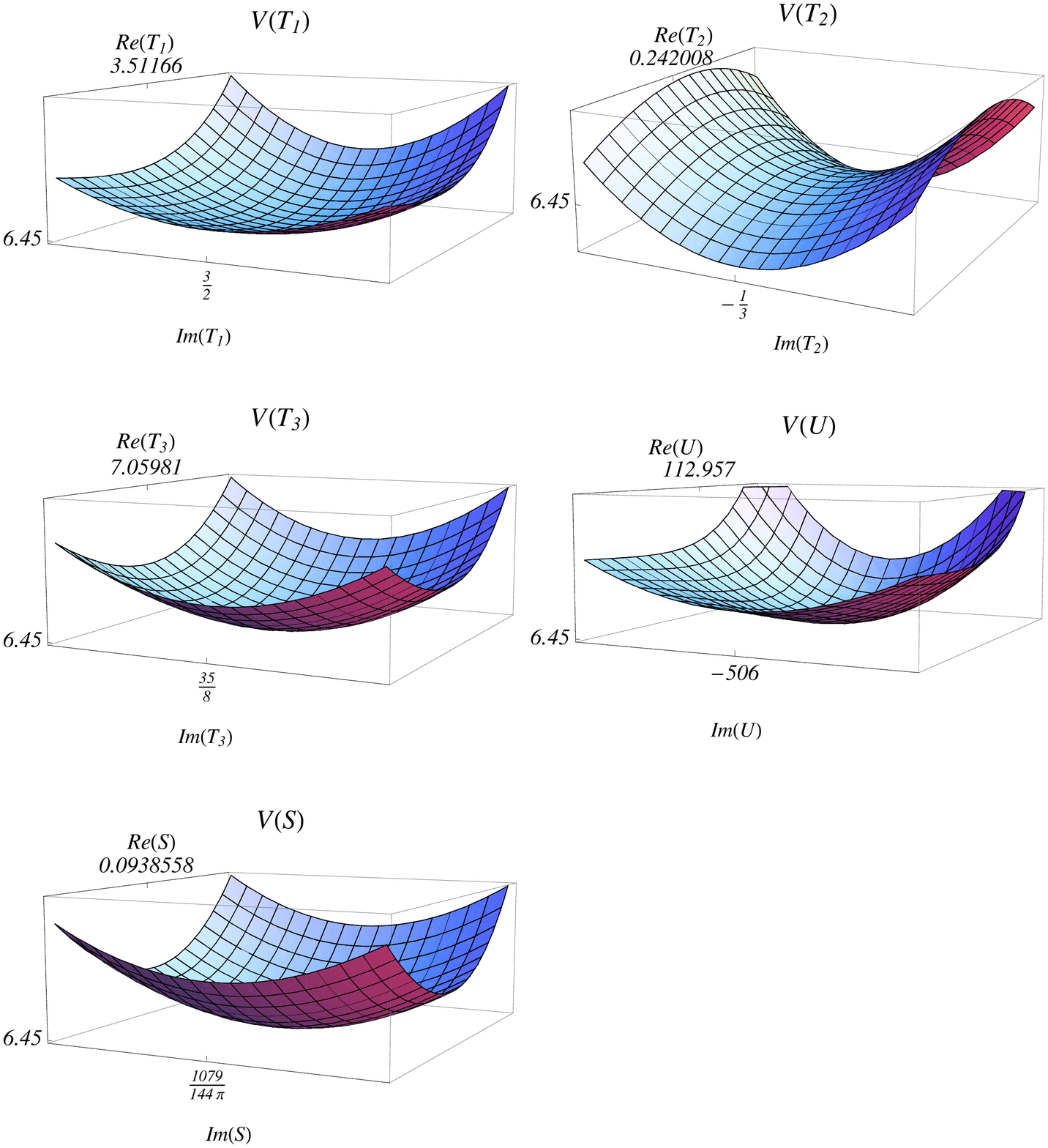}
\vskip-15mm
\caption{\label{fig:Vacuum} Scalar potential as a function of the bulk moduli of an orbifold MSSM candidate 
  for the vacuum presented in tab.~\ref{tab:unstabledSI}. The scalar potential is rescaled by a factor $10^{19}$.}
\end{figure}

Applying eqs.~\eqref{eq:K} and~\eqref{eq:W} to the general expression of the F-term 
potential, $V_F= e^K \left(K^{A\bar{B}}D_AW {D_{\bar B}\overline{W}} - 3|W|^2 \right)$
(the indices $A,B$ label all chiral supermultiplets present and $D_A W = \partial_A W +  W\,\partial_A K$),
we have computed the scalar potential of the model. In the limits of our assumptions
and with $c=1/10$, $20A_1=100A_2=A_3=1/10$, $M_d=1/65$,
we have found numerically dozens of vacua out of which we chose the one shown in 
tab.~\ref{tab:unstabledSI}.  The scalar potential is plotted in fig.~\ref{fig:Vacuum}. Unfortunately, 
all vacua are unstable for they turn out to have at least one tachyon. In the
current model, the unstable direction is dominated by Re$T_2$. 
The problem relies most likely in the fact that there are plenty of vacua.~\footnote{It could also be related
to the no-scale-like structure of our K\"ahler potential and/or the direction of supersymmetry breakdown,
as suggested in~\cite{Covi:2008ea,Covi:2008zu}, but the tools to address analytically these questions in our system
must still be developed.} Exploring all of
them becomes rapidly a numerical issue because of the complicated structure of the potential of
orbifold models. There are, however, reasons to expect that a stable solution will emerge in a
more refined search that will be presented elsewhere~\cite{Akrami:2011xx}. 

On the bright side, the vacua we find have the following properties:
\begin{itemize}
\item all vacua are de Sitter with small cosmological constant;
\item our setting favors anisotropic  compactifications, which
is one  alternative to achieve precision gauge coupling unification and to solve
the infamous hierarchy between the string scale and the unification scale $M_{GUT}$;
\item the overall volume
turns out to be always large, in consistency with the K\"ahler approximation;
\item although difficult, it is possible to reach larger values of $S$, close to
the expected value to accomplish gauge coupling unification at $M_{GUT}$.
\end{itemize}
We expect these features to be generic consequences of moduli stabilization via gaugino 
condensation in heterotic orbifolds.

\section{Some speculation: cosmological prospects}
\label{sec:cosmology}

If a metastable vacuum can be found in these models, there are further questions connected
to the dynamics of the moduli that must be answered.
A successful string model should be able not only to reproduce the particle physics
of our universe, but also to explain the structure and dynamics of the cosmos. One
persisting problem is to explain the origin and dynamics of the rapid period of
expansion of the observable dimensions, known as inflation (for a nice review, see~\cite{Mazumdar:2010sa}). 
Typically, it is assumed
that there is some field, the inflaton, whose potential is almost flat and that has a minimum somewhere. 
Many single and multi-field inflationary models have been concocted and 
lead to acceptable e-folds, primordial fluctuations and nongaussianities.

In the context of string theory, there have been efforts mostly in the type IIB 
string~\cite{Kallosh:2004yh,Lalak:2005hr,Covi:2008cn,Conlon:2008cj}
to derive inflation from first principles. Remarkably, it has been found that, in certain scenarios,
admissible fluctuations and nongaussianities are more natural in a universe dominated by multi-field
dynamics~\cite{Burgess:2010bz}, as is always the case in string constructions. However,
the heterotic string has not been explored sufficiently in this regard. 

\begin{figure}[t]
  \begin{center}
    \includegraphics[width=0.25\textwidth]{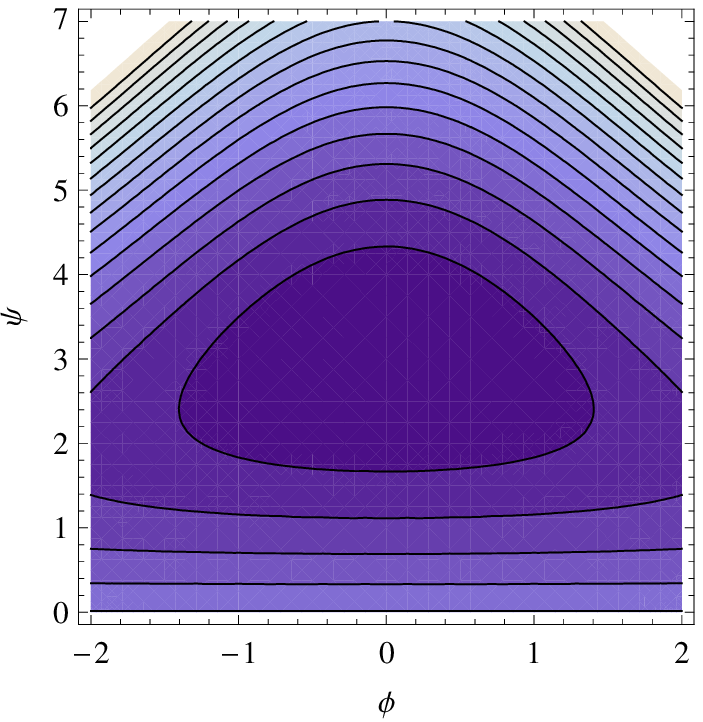}
    \hskip9mm
    \includegraphics[width=0.37\textwidth]{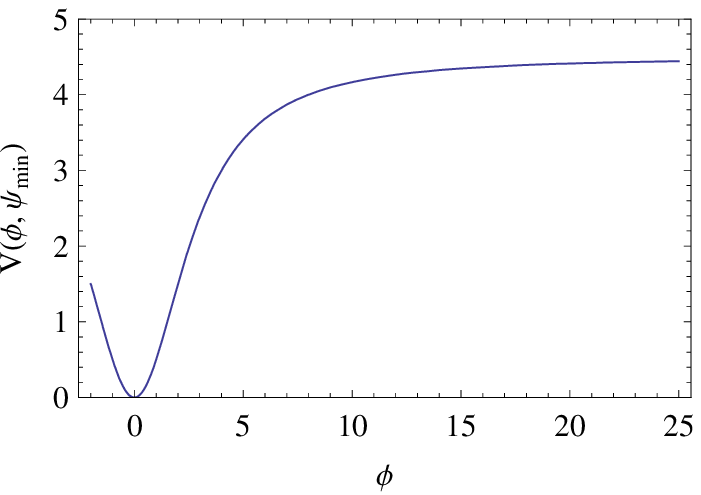}
  \end{center}
  \caption{\label{fig:inflaton} Modified hybrid inflation. The field $\psi$ develops quickly a VEV 
     inducing an inflationary potential for $\phi$. We have taken $V=(\psi-3)^2/2+\phi^2\psi^2/16$.}
\end{figure}

In the particular scenarios we study here, we find generically some scalar fields
whose squared masses lie between $10^{-100}$ and $10^{-30}$ in Planck units, look e.g. at the masses of
Im$T_3$ and Im$U$ (more precisely, Im$\rho_3$ and Im$\rho_U$) in tab.~\ref{tab:unstabledSI}.
Since these quantities correspond to the curvature of the potential on the directions of those fields
in moduli space, there are some almost flat directions which, nevertheless, lead to a minimum. The left panel
of fig.~\ref{fig:inflaton} is a cartoon of this situation involving only two fields, where we have denoted 
them by $\phi$ and $\psi$. 

We can now consider two different situations: when the coupling between $\phi$ and $\psi$ cannot
be neglected, and when $V(\phi,\psi)\approx V(\phi)+V(\psi)$. In the former case, we might arrive at
the scenario described by hybrid inflation models. If the field $\psi$ rapidly develops a VEV, the 
resulting scalar potential $V(\phi)=V(\phi,\psi_\text{min})$ is just the ideal slow-roll potential,
as displayed in the right panel of fig.~\ref{fig:inflaton}. In the latter case, the fields 
with quasi-flat potential might be suitable to accommodate the multi-field curvaton-inflaton scenario
investigated in~\cite{Burgess:2010bz}, yielding more accessible values for cosmological
observables, such as the nonlinearity parameter $f_{NL}$.
Of course, whether or not we get such behaviors in actual orbifold models is still a question we shall
analyze elsewhere.
It seems nevertheless plausible to combine the experience gathered in type IIB and the rich 
structure of heterotic orbifolds to arrive at a promising stringy inflationary scenario, goal
that now can and must be pursued.

\section*{Acknowledgments}
I would like to thank S. Parameswaran and I. Zavala for a very pleasant and fruitful collaboration, and
W. Buchm\"uller and S. Raby for very useful discussions. I am also thankful to the organizers of
the XIV Mexican School on Particles and Fields for their invitation. In particular, 
I am indebted to A. G\"uijosa, O. Loaiza-Brito and A. Raya for their support. 

\section*{References}
\providecommand{\newblock}{}

\end{document}